\documentclass{article}

\RequirePackage{mathptmx} 
\usepackage{PRIMEarxiv}
\usepackage{amsmath}
\usepackage{cite}
\usepackage{amsmath,amssymb,amsfonts}
\usepackage{algorithmic}
\usepackage{graphicx}
\usepackage{siunitx}
\usepackage{textcomp}
\usepackage{upgreek}
\usepackage[utf8]{inputenc} 
\usepackage[T1]{fontenc}    
\usepackage{hyperref}       
\usepackage{url}            
\usepackage{booktabs}       
\usepackage{amsfonts}       
\usepackage{nicefrac}       
\usepackage{microtype}      
\usepackage{lipsum}
\usepackage{fancyhdr}       
\usepackage{graphicx}       
\graphicspath{{media/}}     

\title{Correcting directional dark-field x-ray imaging artefacts using position-dependent image deblurring and attenuation removal}

\author{
 Michelle K Croughan\textsuperscript{1}, David M Paganin\textsuperscript{1}, Samantha J Alloo\textsuperscript{1,2}, Jannis N Ahlers\textsuperscript{1}, \\
 \bf{Ying Ying How\textsuperscript{1}, Stephanie A Harker\textsuperscript{3},  Kaye S. Morgan\textsuperscript{1} }\\
 \\
\textsuperscript{1}
Monash University, School of Physics and
Astronomy, Clayton, VIC 3800, Australia\\
\textsuperscript{2}
University of Canterbury, School of Physical and Chemical Sciences, Christchurch, 8041, New Zealand
\\
\textsuperscript{3}
Monash University, School of Medicine, Melbourne, 3800, Australia
\\
\texttt{email: Michelle.Croughan@monash.edu}}

\begin{document}
\maketitle

\begin{abstract}
In recent years, a novel x-ray imaging modality has emerged that reveals unresolved sample microstructure via a ``dark-field image'', which provides complementary information to conventional ``bright-field'' images, such as attenuation and phase-contrast modalities. This x-ray dark-field signal is produced by unresolved microstructures scattering the x-ray beam resulting in localised image blur. Dark-field retrieval techniques extract this blur to reconstruct a dark-field image. Unfortunately, the presence of non-dark-field blur such as source-size blur or the detector point-spread-function can affect the dark-field retrieval as they also blur the experimental image. In addition, dark-field images can be degraded by the artefacts induced by large intensity gradients from attenuation and propagation-based phase contrast, particularly around sample edges. By measuring any non-dark-field blurring across the image plane and removing it from experimental images, as well as removing attenuation and propagation-based phase contrast, we show that a directional dark-field image can be retrieved with fewer artefacts and more consistent quantitative measures. We present the details of these corrections and provide ``before and after'' directional dark-field images of samples imaged at a synchrotron source. This paper utilises single-grid directional dark-field imaging, but these corrections have the potential to be broadly applied to other x-ray imaging techniques.
\end{abstract}


\section*{Introduction}
X-ray analogues of visible light dark-field microscopy~\cite{gageModernDarkFieldMicroscopy1920,gaoDarkFieldMicroscopyRecent2021} provide sample information that is complementary to existing x-ray imaging techniques. Some dark-field imaging work directly images the x-rays scattered by the sample separately from the direct x-ray beam~\cite{davisPhasecontrastImagingWeakly1995,simonsDarkfieldXrayMicroscopy2015}. Note, this type of dark-field image contains scattered x-rays from resolved and unresolved sample structure, unlike the work presented here. In this paper we discuss x-ray dark-field scattering that may be described as a ``diffusive dark-field'' signal, which is generated by unresolved spatially-random sample microstructure. This diffusive dark-field signal is henceforth referred to as just the ``dark-field'' and it is this diffusive signal that is the focus of this paper, together with most of the references within.

Due to the x-ray dark-field signal's sensitivity to unresolved sample microstructure it can be utilised to capture local statistical information about this microstructure. Collecting x-ray images of sample microstructures is typically limited by the spatial resolution of the imaging system. Fortunately, microstructures smaller than the pixel size can scatter the x-ray beam and create a measurable dark-field signal~\cite{Ando_2001,pagot2003,pfeiffer_hard-x-ray_2008} that can be captured by a lower resolution imaging system. Additionally, aligned elongated microstructures can create a directional dark-field signal where the scattering is greater in the direction perpendicular to the long axis of the microstructures, encoding this directional information in the image~\cite{ Jensen_2010_A,jensenDirectionalXrayDarkfield2010_B}. Measuring this dark-field signal reveals the presence of microstructures unresolved by the imaging system, and can potentially also provide quantitative measurements of these microstructures~\cite{Prade_2015,taphorn2020grating, How:22, How_2023}. Dark-field x-ray imaging is already being applied in settings that can benefit this, such as clinical and pre-clinical trials to improve the diagnostic power of chest radiography~\cite{urbanDarkFieldChestRadiography2023} and breast cancer screening~\cite{emons2020,aminzadeh2022imaging}.

Various techniques exist to capture and retrieve the x-ray dark field~\cite{pfeiffer_hard-x-ray_2008,Jensen_2010_A,zdora_x-ray_2017,How:22,pagot2003,rigon2007_abi_3,endrizzi2014}. This paper utilises single-grid imaging~\cite{morgan2011quantitative,wen2008} adapted for directional dark-field retrieval~\cite{croughanDirectionalDarkfieldRetrieval2023}. This technique uses an absorption or phase grid as a reference pattern to produce vertical and horizontal intensity variations, similar to adjacent techniques that directly resolve the intensity patterns created by a single grating or speckle reference pattern~\cite{kagias2DOmnidirectionalHardXRayScattering2016,Pavlov2021,Smith2022}. For single-grid imaging, a reference pattern image is collected called the reference-only image. Then a sample is placed in the x-ray beam with the reference pattern and the sample-and-reference image is collected. The attenuation of the x-ray beam by the sample results in a reduction of intensity in the sample-and-reference image. The phase properties of the sample will cause the refraction of the beam resulting in transverse displacement of intensity as the beam propagates to the detector, known as propagation-based phase contrast~\cite{snigirevPossibilitiesRayPhase1995,cloetensPhaseObjectsSynchrotron1996}. This type of contrast is often utilised to help highlight the boundaries of weakly attenuating materials as it results in bright and dark fringes, known as phase fringes, around the edges of the sample. Finally, if the unresolved sample microstructure scatters the x-rays, this will locally blur the sample-and-reference image, reducing the local visibility of the reference pattern. These changes to the reference pattern can be measured and attributed to the sample attenuation, phase, and dark-field signals. 

\begin{figure}[ht!]
\centering
\includegraphics[width=\linewidth]{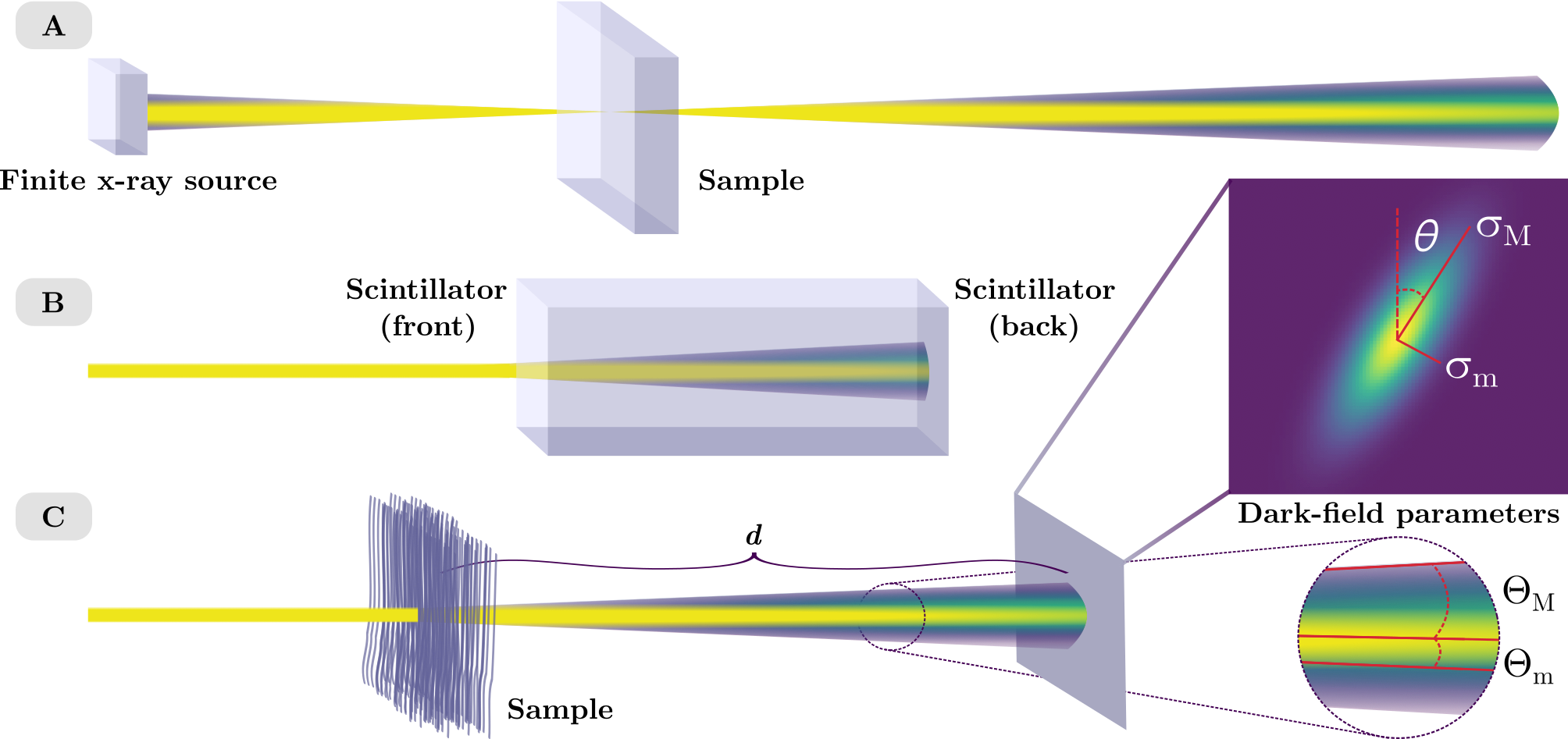}
\caption{Visualisations for source-size blur (A) where a single point in the sample will have x-rays incident on it at slightly different angles due to the finite x-ray source size, resulting in a blurring of the sample image as the beam propagates downstream of the sample. An example of detector point-spread-function blur (B), where the scintillator absorbs the x-ray light and emits a cone of visible light, with the camera capturing the diffused intensity distribution seen at the back of the scintillator. Model for directional dark-field blur (C) where unresolved microstructures in the sample scatter the x-ray beam. The directional dark-field blur is modelled by an asymmetric Gaussian with the blurring widths described by the semi-major and semi-minor standard deviations $\sigma_M$ and $\sigma_m$, respectively, and the dominant scattering angle $\theta$, defined using a clockwise rotation from the vertical axis. The divergence of the beam due to the dark field is described by scattering angles given by $\Theta_M$ and $\Theta_m$.}
\label{Fig: blur model}
\end{figure}

We identified two issues with the quality of the dark-field signal retrieved using single-grid x-ray imaging. Firstly, the scattering angles are intrinsic to the sample, and thus the measured scattering angle should be the same regardless of the sample-to-detector distance. In other words, the associated blurring widths at the image plane should increase proportionally with the sample-to-detector distance. However, it has been observed that the measured scattering angle from a sample can vary with the sample-to-detector distance~\cite{croughanDirectionalDarkfieldRetrieval2023}. Secondly, artefacts appear in dark-field images around the edges of sample features
~\cite{croughanDirectionalDarkfieldRetrieval2023}. We hypothesised that any experimental factors that modify the pattern visibility in an image, or affect our ability to correctly measure the visibility, can reduce the quality of the retrieved dark-field image, where a reduction in quality could be seen as either the presence of artefacts or as inaccurate quantitative measurements. For example, a finite x-ray source size or optical elements such as scintillators and lenses can introduce blurring in the image that does not originate from the sample dark-field signal; such sample-independent blur will be referred to in this paper as ``non-dark-field blur''. A finely-structured illumination, like that introduced in a single-grid set-up in order to be sensitive to dark-field blurring, is hence also very sensitive to any such non-dark-field blurring, in a way that conventional attenuation-based x-ray imaging is not. Some examples of non-dark-field blur are shown in Figure~\ref{Fig: blur model}~A and~B alongside a Gaussian model for sample induced directional dark-field blur shown in Figure~\ref{Fig: blur model}~C. The artefacts that are present around sample features in the dark-field images appear to be anywhere that the sample creates intensity variations either from spatially-varying absorption or propagation-based phase-contrast fringes. The sample attenuation and propagation-based phase contrast can make it difficult to differentiate the dark-field signal from other information present in the image, particularly where there are large intensity gradients such as phase fringes at the edges of the sample. Propagation-based phase-contrast fringes have been shown to cause edge artefacts in phase retrieved images using speckle-based imaging~\cite{morganXrayPhaseImaging2012}, and removal of the attenuation and propagation-based phase contrast from the experimental image prior to phase retrieval was successful in reducing these artefacts~\cite{wangSpeckletrackingXrayPhasecontrast2017}. We expect single-grid dark-field imaging will also benefit from this correction.

In this paper, we present an image correction technique to isolate the blurring from dark-field effects prior to dark-field retrieval. This improves the appearance and the quantitative results of the directional dark-field images retrieved from single-grid x-ray image data. Most x-ray dark-field imaging techniques measure a reduction in visibility of intensity fluctuations in an image(s) in comparison to a reference image(s)~\cite{pfeiffer_hard-x-ray_2008, zdora_x-ray_2017,How:22}. So while these improvements are demonstrated with single-grid imaging, we believe that this image correction technique can be broadly applied to improve other dark-field imaging techniques, beginning with adjacent techniques such as speckle-based imaging. We begin by describing the forward problem with the introduction of a position-dependent blurring operator to model the non-dark-field blurring, and discuss in more detail the relationship between the experiment and the directional dark-field image quality issues highlighted above. We then present an iterative deblurring algorithm to remove the non-dark-field blur from the experimental images. Finally, we show that these correction techniques improve the directional dark-field images with experimental data of two different samples.

\section*{The forward problem: modelling the x-ray image}
Non-dark-field blurring introduced by the experimental setup can vary with position in the image plane. For example, a scintillator with fluctuating thickness or damage can produce spatially varying point-spread-function blurring and optical elements can produce spherical aberration, where there is blurring towards the edges of the image while the centre of the image appears sharp. To model this blurring we introduce a position-dependent blurring operator, denoted by $\mathcal{B}$. This integral operator is applied to an image $I(x,y)$ to map it to a blurred image $I'(x,y)$. We define the operator and operator notation as
\begin{equation}
    I'(x,y) =\int^{b}_{a}\int^{b}_{a} \kappa ((x',y'),(x-x',y-y')) I(x',y') dx' dy'  \quad \Longrightarrow \quad I'= \mathcal{B} I,
    \label{Eq: blurring operator definition}
\end{equation}
where $\kappa$ contains the blurring kernel for each position in the unblurred image. These kernels describe how the intensity at a position $(x,y)$ in $I$ is blurred, with the result that the same position in $I'$ has both lost intensity that has been blurred into other regions and gained intensity from neighbouring regions. The limits of integration $a$ and $b$ only need to be considered in the domain relevant for accurately mapping a finite-sized unblurred image to a blurred image of the same size. For the application in mind, a normalised asymmetrical 2D Pearson VII function is used to describe the blurring kernels, which we derived to be
\begin{equation}
    PVII(i, j) = \frac{m-1}{\pi \alpha_m \alpha_M} \left( 1 + \left(R \begin{bmatrix}
        i\\
        j
    \end{bmatrix}
    \right)^T \begin{bmatrix}
        \alpha^{-2}_m & 0 \\
        0 & \alpha^{-2}_M
    \end{bmatrix}
    R \begin{bmatrix}
        i\\
        j
    \end{bmatrix}\right)^{-m},\quad R = \begin{bmatrix}
    \cos \phi & -\sin \phi\\
    \sin \phi & \cos \phi
    \end{bmatrix}, \quad m > 1,
    \label{Eq: PVII definition}
\end{equation}
where $\alpha_M$ and $\alpha_m$ are the parameters that describe the blurring width of the function in the semi-major and semi-minor directions, $R$ is the rotation matrix, $T$ is the matrix transpose operator, $\phi$ determines the rotation in a clockwise direction from the vertical axis, and $m>1$ is a parameter that alters the shape of the function to go between a Lorentzian ($m\longrightarrow1$) and a Gaussian distribution ($m\longrightarrow \infty$). Parameters $m, \phi, \alpha_M, \alpha_m$ are varied to produce the blurring kernel required for a particular position $(x,y)$ in $I$, and $i$ and $j$ represent the difference in location $(i,j)=(x-x',y-y')$. Lorentzian distributions can be good models for some types of blurring kernels as they contain long tails that do not approach zero as rapidly as Gaussian distributions~\cite{husseyImprovingQuantitativeNeutron2013}. Using Equation~\ref{Eq: PVII definition} to model the non-dark-field blurring is powerful as it allows for models using both types of distributions. 

We approximate the experimental reference-only image $I_r'$ as the result of taking the unblurred image of the reference pattern $I_r$ and applying the non-dark-field blur
\begin{equation}
    I_r' = \mathcal{B}_{N} I_r,
    \label{Eq: reference only image}
\end{equation}
where $\mathcal{B}_{N}$ is the position-dependent blurring operator for the non-dark-field blur, such as the source-size blur and/or the detector point-spread-function. Depending on the experiment one could consider other potential sources of non-dark-field blur and even split these sources into separate blurring operations if required.
The unblurred sample-only image $I_s$ is the combination of the sample attenuation and propagation-based phase contrast, seen when only the sample is in the x-ray beam, excluding both the dark-field and non-dark-field image blur. Another position-dependent blurring operator $\mathcal{B}_{DF}$ is included to model the sample dark-field blur. By assuming that the non-dark-field blurring is slowly varying and that the dark-field blurring is localised we approximate the experimental sample-only image as
\begin{equation}
    I_s' = \mathcal{B}_{N}\mathcal{B}_{DF} I_s.
    \label{Eq: sample only image}
\end{equation}
When the sample and reference pattern are placed in the x-ray beam together, the unblurred image $I_{sr}$ describes the combined attenuation and propagation-based phase contrast seen when both objects are illuminated. The experimental sample-and-reference image will then be modelled by
\begin{equation}
    I_{sr}' = \mathcal{B}_{N} \mathcal{B}_{DF}I_{sr},
    \label{Eq: sample-and-reference image}
\end{equation}
where the non-dark-field blur and dark-field blur affect both the image of the sample and the reference pattern.

The non-dark-field blur is present in both the reference-only and sample-and-reference images. However, due to the absorption of the x-ray beam by the sample, the amount of intensity displaced by this non-dark-field blur at each pixel can be different in the sample-and-reference image compared to the reference-only image. The detector point-spread-function in particular can affect the image over a large scale, and so a change in intensity incident on one region of the image can have a non-negligible effect on the intensity recorded on the other side of the image~\cite{husseyImprovingQuantitativeNeutron2013}. 
The sample attenuating the x-ray beam can alter the amount of intensity distributed by the non-dark-field blur and affect the visibility of the reference pattern locally as well as in distant parts of image. This means the visibility of the reference pattern can be changed purely due to attenuation of the x-ray beam, making quantitative dark-field retrieval difficult. Non-dark-field blur can also result in the reference pattern being more visible in some directions than others. This can result in dark-field artefacts if the less-visible direction approaches dark-field saturation, where the reference pattern features become close to or completely blurred out. Attenuation and propagation-based phase contrast can also affect the dark-field retrieval by creating large intensity gradients in an image, in a similar manner to how phase-contrast fringes can affect phase retrieval in speckle-based imaging~\cite{wangSpeckletrackingXrayPhasecontrast2017}. For example, a sudden change in intensity over the length scale of the reference pattern feature size could appear as a change in visibility rather than a change in mean intensity. These effects can make it difficult for dark-field retrieval algorithms to determine if a change between the reference-only and sample-and-reference data is due to absorption, phase, or dark-field contrast.

\section*{The backward problem: isolating the dark field}
In order to improve the dark-field retrieval we propose that it could be beneficial to compare the images $I_r$ and $\mathcal{B}_{DF}I_r$, so the only difference between the two images is due to dark-field effects. These images can be computed using the model we have created for the experimental reference-only image (Equation~\ref{Eq: reference only image}) and sample-and-reference image (Equation~\ref{Eq: sample-and-reference image}). This can be done by removing the non-dark-field blur using a position-dependent deblurring operator, which we derive below, and then removing the attenuation and propagation-based phase contrast due to the sample by dividing out a sample-only image. An alternate approach to account for non-dark-field blur has been to incorporate the combined affects of non-dark-field and dark-field blur in an implicit model~\cite{paganinParaxialDiffusionfieldRetrieval2023}. However, in this paper we present a correction method that can be applied directly to experimental images for the use with explicit retrieval techniques.

In order to deblur an image the inverse of the blurring operator described by Equation~\ref{Eq: blurring operator definition} is required. As digital x-ray images are discrete, Equation~\ref{Eq: blurring operator definition} can be written as an $N \times N$ matrix operation, where $N$ is the number of pixels in the x-ray image. 
The inverse matrix exists, however solving it directly is not desirable for two reasons. Firstly, inverting an $N \times N$ matrix is not computationally feasible for standard x-ray image sizes (where $N \gtrsim 10^6$) even with cluster-based computing resources. Secondly, while the blurring kernel can vary from pixel to pixel, nearby pixels may have very similar blurring kernels, which would result in this matrix being close to singular (where a singular matrix is non-invertible) and so inverting it is likely to lead to strong numerical instabilities. To overcome this we combine a pseudoinverse operator with a Richardson-Lucy~\cite{Richardson1972,Lucy1974} style algorithm to iteratively deblur the experimental images. 

We derive a suitable pseudoinverse by starting with the operator notation of Equation~\ref{Eq: blurring operator definition} and include two identity operators `Id' (that leaves the operand unchanged), leaving the equation equivalent to the previous form
\begin{equation}
    (\text{Id} + \mathcal{B} - \text{Id})I = I'.
\end{equation}
Defining that $ \Tilde{\mathcal{B}} = \mathcal{B}-\text{Id}$ and applying the inverse operation $I$ becomes the subject
\begin{equation}
    I = \left(\text{Id} +  \Tilde{\mathcal{B}}\right)^{-1}I'.
    \label{eq: unblurred image}
\end{equation}
The Neumann series can be used to create a series expansion of the inverse operation~\cite{urbanCorrectionXrayScatter2023,FoundationsImageScience}. Adapting Equation~1.213 of the book by Barrett and Myers~(2004)~\cite{FoundationsImageScience} we can define the pseudoinverse operation as
\begin{equation}
    \left(\text{Id} +  \Tilde{\mathcal{B}}\right)^{-1} = \sum_{k=0}^{\infty} \left( -\Tilde{\mathcal{B}}\right)^k = \text{Id} - (\mathcal{B}-\text{Id}) + (\mathcal{B}-\text{Id})^2 - (\mathcal{B}-\text{Id})^3 + ... 
    \label{Eq: pseudoinverse}
\end{equation}
This pseudoinverse operator can deblur images using the original position-dependent blurring operator, avoiding the computationally impractical task of a matrix inversion and providing numerical stability. When applied iteratively to experimental data it was found that setting the maximum value of $k$ to be~1 provided computationally efficient results that were comparable in quality to using a high-order expansion. Thus, we use
\begin{equation}
    \Hat{I} = 2I' - \mathcal{B}I',
    \label{Eq: approx pseudoinverse}
\end{equation}
where $\Hat{I}$ is the approximate solution to $I$ (Equation~\ref{eq: unblurred image}) using only the first and second terms of the Neumann series shown in Equation~\ref{Eq: pseudoinverse}. 

The Richardson-Lucy image restoration algorithm~\cite{Richardson1972,Lucy1974} was adapted to utilise the position-dependent blurring operator (Equation~\ref{Eq: blurring operator definition}) and its pseudoinverse operator (approximated by Equation~\ref{Eq: approx pseudoinverse}) to iteratively deblur, denoting this algorithm with the operator $\mathcal{D}$. We define this algorithm and operator notation as 
\begin{equation}
    \Hat{I}_{n+1} = \Hat{I}_n \left(2\left(\frac{I'}{\mathcal{B}\Hat{I}_n} \right)-\mathcal{B}\left(\frac{I'}{\mathcal{B}\Hat{I}_n} \right)\right)  \quad \Longrightarrow \quad \Hat{I} = \mathcal{D} I',
    \label{Eq: deblurring operator}
\end{equation}
where $\hat{I}_n$ represents the approximate image of $I$ after $n$ iterations. For the first iteration, $\hat{I}_n$ is set to the blurred image $I'$. This algorithm measures how different $I'$ and $\mathcal{B}\hat{I}_n$ are from each other. These differences are then deblurred and used to update $\hat{I}_n$ such that it converges to $I$.

To divide out the sample-induced attenuation and propagation-based phase contrast from the sample-and-reference image, an experimental sample-only image $I_s'$ is required. Depending on the experiment this may be trivial to collect, but potentially may not be possible when imaging a moving sample or a sample that is radiation-dose-sensitive. The single-grid directional dark-field retrieval algorithm~\cite{croughanDirectionalDarkfieldRetrieval2023} reconstructs a sample transmission image, which can be used as an approximate image for $I_s'$. Alternatively $I_s'$ can also be approximated by scaling down and back up the number of pixels in the experimental images. The number of pixels in the experimental reference-only and sample-and-reference images is scaled down by taking the average intensity across each grid-period region in the image. The scaled sample-and-reference image is divided by the scaled reference-only image, and then the number of pixels is increased back to the original size using bilinear interpolation. Lowering the image resolution effectively removes the dark-field blurring of the reference pattern, creating an image that retains only low-spatial-frequency information, which is predominantly due to attenuation. Both of these approximations have lower spatial resolution than a real image of $I_s'$ and may not capture high-spatial frequency information such as phase fringes, yet we show that they can still reduce the appearance of artefacts in x-ray dark-field images that are due to attenuation and propagation-based phase contrast. 

In the context of our experiment, we can use the experimental images (including either a real or approximate sample-only image) and the deblurring operator (Equation~\ref{Eq: deblurring operator}) to approximate $I_r$ and $\mathcal{B}_{DF}I_r$ as
\begin{equation}
    I_r = \mathcal{D}_{N}I_r'
    \label{Eq: 11}
\end{equation}
and
\begin{equation}
    \mathcal{B}_{DF}I_r = \frac{\mathcal{D}_{N} I_{sr}'}{\mathcal{D}_{N} I_s'},
    \label{Eq: 12}
\end{equation}
respectively, where $\mathcal{D}_{N}$ is the deblurring operator for the non-dark-field blur. Typically there is variation in the x-ray beam strength and detector response across the image plane. As Equation~\ref{Eq: 12} is a ratio of two images, these variations largely cancel out and hence the image grey levels describe the intensity as a fraction of the original beam intensity. In order to compare the dark-field-blurred-reference image, $\mathcal{B}_{DF}I_r$ to the reference image $I_r$, it is necessary to perform a flat-field correction on the image resulting from Equation~\ref{Eq: 11} after deblurring (with the deblurred flat-field image). The result will be that the image intensity values are also a fraction of the original beam intensity and the two images can be compared to retrieve the dark field. By isolating the dark field in the sample-and-reference image, using the above correction technique, we show that the artefacts discussed in the introduction will be either removed or less severe in the retrieved directional dark-field images. We also show that this process restores the proportional relationship between measured dark-field blurring width and sample-to-detector distance.

\section*{Experimental data acquisition}
\begin{figure}[ht]
\centering
\includegraphics[width=\linewidth]{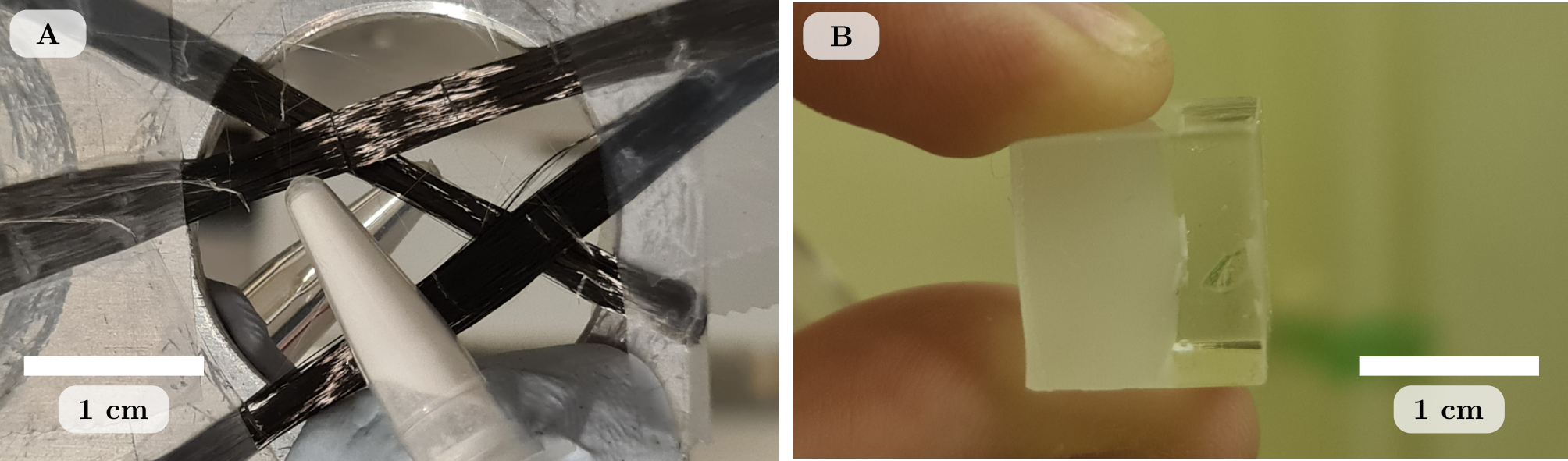}
\caption{Photograph of the dark-field phantom (A) featuring three carbon fibre bundles (typical fibre width of approximately 4 -- 10~$\upmu$m), a tube of water and a tube of 6~$\upmu$m diameter PMMA microspheres. Photograph of the resin sample (B) featuring seashell calcium carbonate flakes and cornflour mixed in with the resin (left portion of resin cube).}
\label{Fig: samples}
\end{figure}

The first data set used to test the described approach contains several objects to create a range of dark-field signals. An aluminium plate with an aperture was used to mount three carbon fibre bundles (typical fibre width of approximately 4 -- 10$~\upmu$m) orientated at different directions and two small polypropylene tubes, one containing water and the other containing 6~$\upmu$m diameter polymethyl methacrylate (PMMA) microspheres. The fibres and tubes were mounted to the plate using sticky tape and adhesive putty. This sample will be referred to as the ``dark-field phantom'' throughout this paper, and a photograph is shown in Figure~\ref{Fig: samples}~A. The carbon fibres and microspheres are expected to produce directional and isotropic dark-field signals, respectively, while the water should produce no dark-field signal. Images of this sample were collected using a monochromatic x-ray beam of 26~keV in hutch 3B of the Imaging and Medical beamline at the Australian Synchrotron. 

The synchrotron x-ray source is larger horizontally than vertically, resulting in strong horizontal source size blurring at large sample-to-detector distances $d$. To investigate the effect of this on the dark-field signal we imaged the dark-field phantom at nine equidistant sample-to-detector distances between $d =$~0.5~m and $d =$~4.5~m, with a 0.5~m separation between each distance. An absorption grid with 90~$\upmu$m holes and 65~$\upmu$m thick grid lines (Essa ISO 3310 ASTM stainless steel geological sieve) was used to create the reference pattern. This grid was placed 0.4~m upstream of the sample, which was the closest possible distance it could be placed to the sample given the physical experimental set-up. Sample-only, reference-only, and sample-and-reference data sets were collected. For each of these data sets 30~exposures of 0.25~s each were collected to be averaged together to improve the signal-to-noise ratio. The ``Ruby'' detector was used, a lens-coupled scintillator (25~$\upmu$m Gadox) with an sCMOS sensor producing an effective pixel size of 21.07~$\upmu$m to 20.47~$\upmu$m for 0.5~m to 4.5~m sample-to-detector distance. Flat-field and dark-current images were also collected to correct for intensity variations in the beam and detector electronic noise, respectively. 

The absorption slits located in hutch 3B were used to narrow the x-ray beam down to less than one pixel in size. This was used to measure the combination of the source-size blur and detector point-spread-function at nine different positions on the detector (each corner, edge, and the centre). This was done at slit-to-detector distances of 0.9~m and 7.65~m (the minimum and maximum distances available). 100~exposures at 2~s each were collected for each blurring kernel at both distances.

The second sample is made of seashell calcium carbonate flakes embedded in an epoxy resin cube. The left half of the cube had cornflour mixed in prior to the resin setting. This sample was selected for its potential to produce both a relatively uniform dark-field signal from the cornflour and an unknown dark field from the seashell flakes, with a photograph shown in Figure~\ref{Fig: samples}~B. Images were collected using a monochromatic x-ray beam of 25~keV in hutch~2 of the 20XU beamline at the SPring-8 Synchrotron, Japan.

The resin sample was imaged at a sample-to-detector distance of 36.5~cm and the same phase grid used in Morgan~\textit{et~al.}~(2013)~\cite{morganSensitiveXrayPhase2013} was placed 14~cm upstream of the sample to create a grid reference pattern with a 2.7~$\upmu$m period. We collected 100~images with an exposure of 0.1~s for the reference-only and sample-and-reference data. Flat-field and dark-current images were collected. The 20XU Beam Monitor type III was used, which consists of an ORCA flash4.0 CMOS sensor coupled with an optical lens and crystal scintillator providing an effective pixel size of 0.49~$\upmu$m. This detector set up distorts the x-ray images by introducing radial blurring from the optical lens which becomes stronger towards the edges of the field of view. We investigate the effect this lens blurring has on the directional dark-field images.

\section*{Data processing and results}
\subsection*{Directional dark-field retrieval with uncorrected and corrected data}
The non-dark-field blurring kernels at the Imaging and Medical beamline of the Australian Synchrotron were captured by collecting images of a sub-pixel x-ray beam incident on nine different positions across the detector, at 0.95~m and 7.65~m slits-to-detector distance. Each image set was averaged to reduce the image noise and the dark-current image was subtracted to remove the detector's electronic noise. Equation~\ref{Eq: PVII definition} was fit to each kernel, thereby providing values for $m$, $\phi$, $\alpha_M$, and $\alpha_m$. These varied across the image plane, for example at the 0.95~m distance the centre of the image had kernel parameters of $m = 1.33$, $\phi = 0.00$~rad, $\alpha_M = 1.99$~pixels, and $\alpha_m = 1.73$~pixels, and the top left region of the image had $m = 1.50$, $\phi = 0.54$~rad, $\alpha_M = 3.53$~pixels, and $\alpha_m = 2.07$~pixels. 
These results were interpolated using the Clough Tocher 2D interpolation~\cite{ALFELD1984} to approximate the kernel parameters for each pixel in the image. The parameters for the 0.95~m and 7.65~m non-dark-field blurring kernels were linearly interpolated to compute the estimated kernels at each of the distances the dark-field phantom was imaged at. 

\begin{figure}[ht!]
\centering
\includegraphics[width=\linewidth]{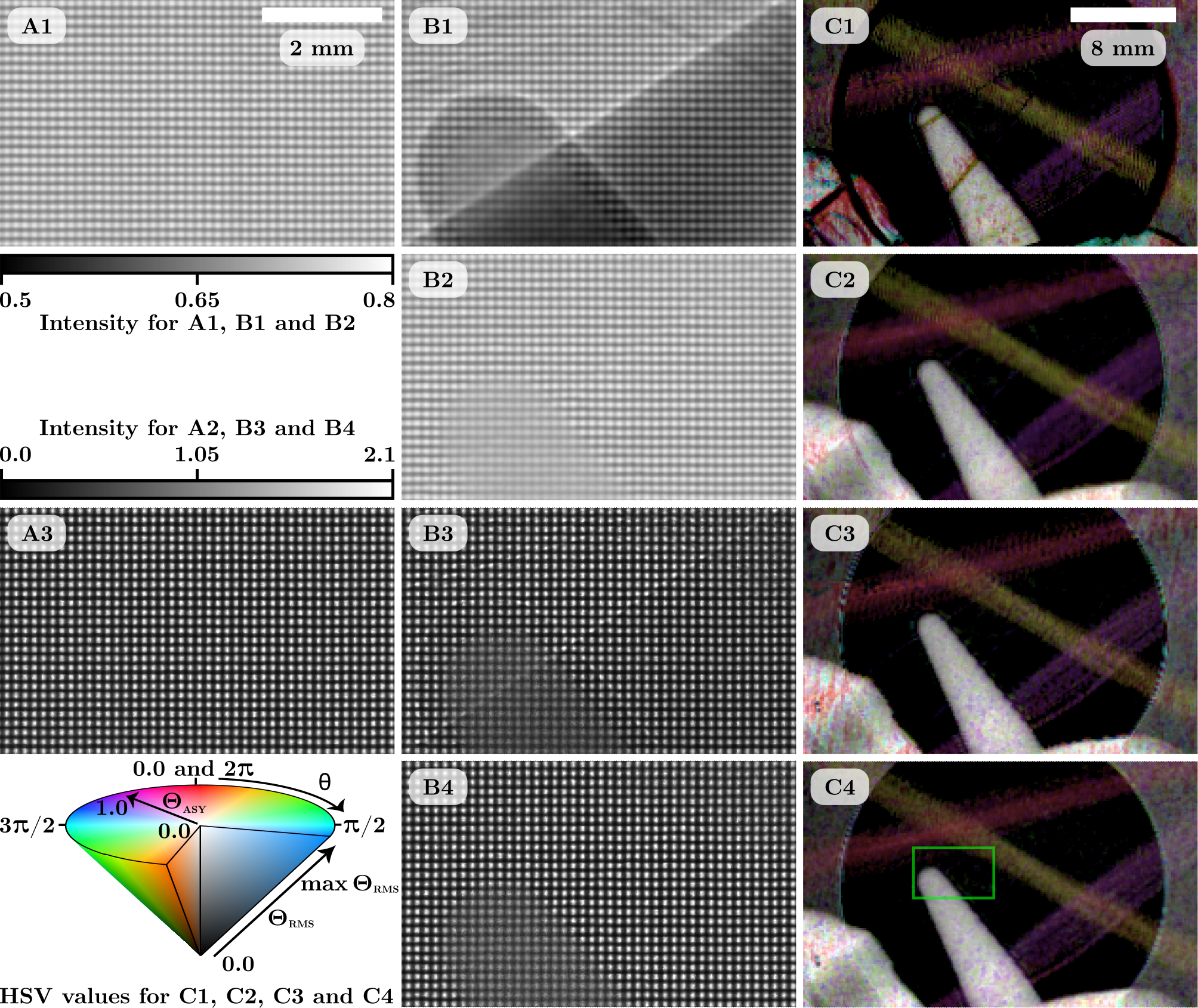}
\caption{The directional dark-field images of the dark-field phantom at sample-to-detector distance of 3.5~m with various corrections applied prior to retrieval. Sections of the reference-only and sample-and-reference images are shown in columns A and B respectively, with the region shown highlighted in~C4 by the green box. A1~and~B1 show the uncorrected image data. They were used to retrieve the directional dark-field image shown in~C1. B2~is the sample-and-reference image divided by the sample-only image, it is used with~A1 to retrieve~C2. A3~and~B3 are the deblurred reference-only and sample-and-reference images used to retrieve~C3. B4~is the deblurred sample-and-reference image divided by the deblurred sample-only image, which is used with~A3 to retrieve~C4. All of the directional dark-field images have the brightness scaled between zero and the maximum dark-field strength $\Theta_{RMS}$ of 13.8~$\upmu$rad. }
\label{Fig: corrections comparions}
\end{figure}

At each sample-to-detector distance, we collected sample-only, reference-only, and sample-and-reference images of the dark-field phantom. Each image set was averaged to reduce noise and had the dark-current image subtracted from it. After this, the images were processed in four different ways prior to dark-field retrieval. Magnified sections of a small region of the images are shown in Figure~\ref{Fig: corrections comparions}: the images in~A1 and~B1 have been flat-field corrected but have had no other processing done, hence they will be referred to as the uncorrected data. B2~is the sample-and-reference image shown in~B1 divided by the sample-only image to remove the attenuation and propagation-based phase contrast. A3~and~B3 are the images deblurred using Equation~\ref{Eq: deblurring operator}, with non-dark-field blurring kernels of $100 \times 100$ pixels in size for 30~iterations. The images were then flat-field corrected after deblurring. B4~is the deblurred sample-and-reference image divided by the deblurred sample-only image. This gives four data sets for comparison: uncorrected, $I_s'$ removed only, deblurred only, and deblurred-and-$\mathcal{D}_NI_s'$ removed. Due to the source-size blur the vertical grid lines are less visible compared to the horizontal grid lines in the uncorrected images. In comparison, Figure~\ref{Fig: corrections comparions}~A3 and~B3 show the improvement in the visibility of the horizontal and vertical grid lines after each image is deblurred. Figure~\ref{Fig: corrections comparions}~B2~(uncorrected) and~B4~(deblurred) show the sample-and-reference images that have had the sample-only image divided out. This process is successful at removing the attenuation and propagation-based phase contrast. Note the different greyscale bars for the images with and without deblurring. We hypothesise that the grey values greater than 1 in the deblurred images could be due to a combination of reasons. The x-ray beam can diffract around the grid lines of the reference pattern, potentially resulting in greater intensity in the gaps of the grid than seen in the flat-field x-ray beam. Alternatively, an overestimation of the width of the non-dark-field blurring kernels could also cause over-sharpening of the image. Regardless, the dark-field blurring kernels retrieved from the deblurred images remain consistent with those retrieved without correction, likely because both the reference and sample-and-reference images are deblurred in the same way.

The directional dark-field image was retrieved from each set of images using the single-grid directional dark-field retrieval algorithm described by Croughan~\textit{et~al.}~(2023)~\cite{croughanDirectionalDarkfieldRetrieval2023}. To apply this algorithm, all the images were rotated 1.8~degrees clockwise to align the grid lines of the reference pattern to be horizontal and vertical. A Fourier transform of the reference-only image was used to determine the grid period at each distance ranging from 7.26~pixels~($d =$~0.5~m) to 7.52~pixels~($d =$~4.5~m). A local analysis kernel size of $12 \times 12$~pixels was used. The directional dark-field retrieval measures the blurring kernel as an asymmetrical Gaussian with three parameters per grid period: the semi-major and semi-minor blurring widths (standard deviations) $\sigma_M$ and $\sigma_m$, and the dominant scattering direction~$\theta$. These are used to calculate the semi-major and semi-minor scattering angles $\Theta_M$ and $\Theta_m$, and hence determine the dark-field strength~$\Theta_{RMS}$ and asymmetry~$\Theta_{ASY}$~\cite{croughanDirectionalDarkfieldRetrieval2023}. A visualisation for these dark-field parameters is shown in Figure~\ref{Fig: blur model}~C. The dominant scattering direction, asymmetry, and strength are combined to create a qualitative hue-saturation-value image, referred to as the directional dark-field image, which are shown in Figure~\ref{Fig: corrections comparions}~column~C.

Figure~\ref{Fig: corrections comparions}~C1 demonstrates some of the artefacts that occur with single-grid directional dark-field retrieval. These can be seen as edge artefacts around the water tube and aluminium aperture, and some directional dark-field signal from the microspheres, where the signal is expected to be isotropic. Also, the carbon fibres produce phase-contrast fringes across the bundles, resulting in a noisy/textured reconstruction artefact. Figure~\ref{Fig: corrections comparions}~C2,~C3 and~C4 show the directional dark-field images retrieved from the data that has had the $I_s'$ removed only, deblurred only, and deblurred-and-$\mathcal{D}_n I'_s$ removed corrections, respectively. In these images, the artefacts seen in~C1 are either removed completely or significantly diminished. The edges of sample features are no longer producing artefacts, particularly seen where the water tube overlaps with the microspheres and the carbon fibres. The carbon fibre artefacts are particularly reduced in~C2 and~C4 where the sample-only image has been divided out and removed the intensity variations due to the phase-contrast fringes. The appearance of the microspheres is much smoother in the corrected directional dark-field images, with less colour saturation indicating a more isotropic measurement. These image corrections significantly improve the directional dark-field image and also improve the quantitative results for dark-field samples that contain or overlap with strong intensity gradients such as phase-contrast fringes. Weakly scattering samples are required to be imaged at long sample-to-detector distances in order for the dark-field blurring to adequately reduce the reference pattern visibility, however longer distances also increase the magnitude of phase-contrast fringes and certain types of non-dark-field blurring such as source-size blurring. Thus, these corrections can improve results especially for weakly scattering samples where the dark-field signal might be more easily overcome by artefacts.   

\subsection*{Directional dark-field parameters at different sample-to-detector distances}
\begin{figure}[ht!]
\centering
\includegraphics[width=\linewidth]{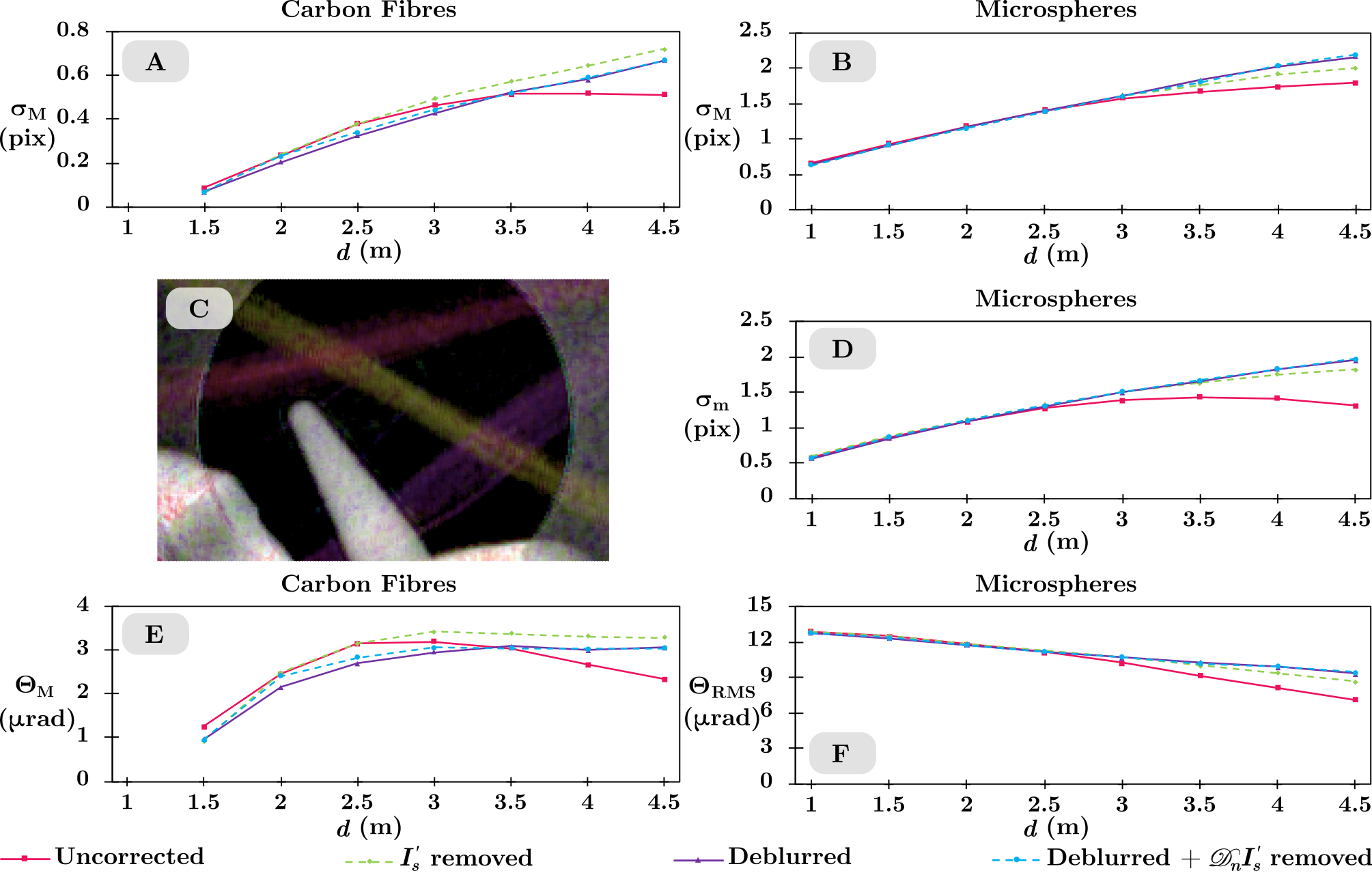}
\caption{Plots showing how dark-field parameters vary with sample-to-detector distance $d$ for uncorrected and corrected data, legend shown at bottom of figure. Semi-major blurring width $\sigma_M$ (A) and scattering angle $\Theta_M$ (E) for carbon fibres. Semi-major blurring width $\sigma_M$ (B), semi-minor blurring width $\sigma_m$ (D), and root-mean-square scattering angle $\Theta_{RMS}$ (F) for microspheres. The data is the mean value of the parameter extracted from the regions shown in C. }
\label{Fig: distances analysis}
\end{figure}

This section describes how the quantitative measurement of the directional dark-field parameters with sample-to-detector distance $d$ is improved with these correction techniques. The blurring widths $\sigma_M$ and $\sigma_m$, scattering angles $\Theta_M$ and $\Theta_m$, and sample-to-detector distance $d$ are all related by~$\sigma = \Theta \times d$~\cite{croughanDirectionalDarkfieldRetrieval2023}. The blurring widths describe how large the dark-field blurring kernel is at the image plane. They are expected to increase proportionally with sample-to-detector distance since the scattering angles should be constant. Small regions of the image containing the microspheres and carbon fibres in the dark-field phantom, , shown in green in Figure~\ref{Fig: distances analysis}~C, were selected to measure the mean semi-major and semi-minor blurring widths at each distance. The shorter sample-to-detector distances showed minimal blurring of the reference pattern due to sub-pixel dark-field blurring kernels, so the 0.5~m and 1~m data are excluded from the carbon-fibre analysis and the 0.5~m data was excluded from the microsphere analysis. The carbon fibres have a very asymmetrical signal, so only the data for the semi-major blurring width and semi-major scattering angle is included.

Figure~\ref{Fig: distances analysis}~A, B, and D show the semi-major blurring width for the carbon fibres and the semi-major and semi-minor blurring widths for the microspheres. The blurring widths measured from the uncorrected data plateaus beyond $d=3$~m, however in theory the blurring width should increase proportionally with $d$. This indicates some mechanism is counteracting the increase in blurring widths expected from the dark-field blurring as $d$ increases. The non-dark-field blurring kernels for the Imaging and Medical beamline were measured to be of a ``near-Lorentzian'' shape (with $1.25 < m < 2.09$ for all kernels) indicating long-range blurring. When the dark-field phantom is introduced, any highly attenuating regions, such as the aluminium mount, will reduce much of the x-ray intensity in the surrounding region due to the significant point-spread-function of the detector. The broad non-dark-field blurring means that the presence of this strong attenuation in one part of the sample-and-reference image can lower the intensity of the rest of the image. This reduction in intensity is seen as a subtraction in grey levels in comparison to the reference-only image and will result in the measured attenuation being higher than expected and the measured dark-field blurring widths being smaller than expected. The source-size blurring becomes stronger at larger values of $d$, resulting in larger non-dark-field blurring kernels. We hypothesise that as $d$ increases, the combined effect of
strong attenuation and non-dark-field blurring described above will become more severe. This will result in the measured dark-field blurring widths not increasing with distance as expected. Correcting the image data by removing the attenuation and propagation-based phase contrast, deblurring the experimental images, or both, will change the dark-field blurring widths measured at longer distances. The intensity redistribution due to the non-dark-field blurring seen in the sample-and-reference image is also seen in the sample-only image; by dividing this out it is possible to correct for the apparent reduction in intensity seen in the non-absorbing regions. By deblurring the experimental images, we are able to return the intensity that was displaced back to its origin in both the reference-only and sample-and-reference images, and so attenuation due to the sample in one region no longer affects the measured attenuation in another region. By doing both we are able to undo the non-dark-field blurring effect whilst also removing the attenuation and propagation-based phase contrast that can cause qualitative artefacts as seen in Figure~\ref{Fig: corrections comparions}. The blurring widths measured from the corrected data continue to increase with distance, indicating the described corrections improve the quantitativeness of the dark-field retrieval, particularly at long sample-to-detector distances.

Figure~\ref{Fig: distances analysis}~E and F show the semi-major scattering angle for the carbon fibres and the root-mean-square scattering angle (dark-field strength) for the microspheres. The scattering angle $\Theta$ is derived from the measured blurring width $\sigma$ and the sample-to-detector distance $d$ and should be a constant. The visibility change $\Delta V$ of the reference pattern is related to the exponent of the negative of the blurring width $\Delta V \propto e^{-\sigma}$~\cite{How_2023}, so there exists an optimal distance where the visibility change between the reference-only and sample-and-reference image is maximised, but not saturated, in relation to the experimental setup~\cite{How_2023}. This means at short distances small blurring widths will correspond to negligible visibility changes that are hard to measure. The measured carbon-fibre scattering angle increases with sample-to-detector distance up to 2.5~m. As the blurring widths are smaller than $1/3$ of a pixel we suspect this is a regime where minimum visibility change is occurring and thus the dark-field blurring kernel is not yet properly measurable. The measured carbon-fibre scattering angle beyond 3~m for the uncorrected data decreases with distance whereas the scattering angle for the corrected data remains constant. The microspheres scatter more strongly than the carbon fibres and so the upward transition seen in the 1.5~m to 2.5~m carbon-fibre $\Theta_M$ data, where the dark-field effects are weak, occurs somewhere between 0~m to 1~m for the microspheres and so is not shown in the plot. The measured $\Theta_{RMS}$ from the microspheres for the uncorrected data has a downward trend with distance rather than remaining constant. The corrected data also has a downward trend but with a lesser gradient. This tells us that while the corrections make the blurring width more consistent with distance, there is an offset or proportionality factor that has not yet been corrected. We suspect that applying the deblurring correction with a larger non-dark-field blurring kernel size will give better improvements beyond what these results already demonstrate, but at the cost of higher computational time. 

Using the corrections to isolate the dark-field blur prior to retrieval has shown that the quantitative measurements of the dark-field signal are more consistent over the distances at which the measurement can be taken. Strongly scattering samples need to be imaged at shorter distances while weakly scattering samples need to be imaged at longer distances. These corrections mean that various samples can be imaged at different distances and the retrieved quantitative results will be more consistent than without correction. 

\subsection*{Attenuation and propagation-based phase contrast correction without an experimental sample-only image}
The improvement in the directional dark-field image in Figure~\ref{Fig: corrections comparions}~C2 compared to Figure~\ref{Fig: corrections comparions}~C1 is the result of dividing out the experimental sample-only image from the sample-and-reference image prior to dark-field retrieval. As shown in Figure~\ref{Fig: corrections comparions}~C3, deblurring the images also reduces many of these artefacts but takes considerably more computational time compared to dividing out the sample-only image. We explored methods of achieving similar improvement where it is not possible to collect an experimental sample-only image $I_s'$. One such scenario is time-sequence imaging, where the sample changes or moves exposure-to-exposure. The first approach is to use the sample transmission image retrieved by the single-grid directional dark-field retrieval algorithm (or any analysis process that retrieves a transmission image) as an approximation to the sample-only image. The second approach is to scale down the number of pixels of the experimental reference-only and sample-and-reference images by a factor of a grid period, divide the reduced sample-and-reference image by the reduced reference-only image, and then scale it back to the original number of pixels. By scaling the images it creates a smoothed approximation of the sample-only image while removing the very local dark-field blurring. These approximate sample-only images for the dark-field phantom were used to divide out the attenuation and propagation-based phase contrast from the sample-and-reference image. The same directional dark-field retrieval process detailed above was used to create the directional dark-field images for our data.

\begin{figure}[ht]
\centering
\includegraphics[width=\linewidth]{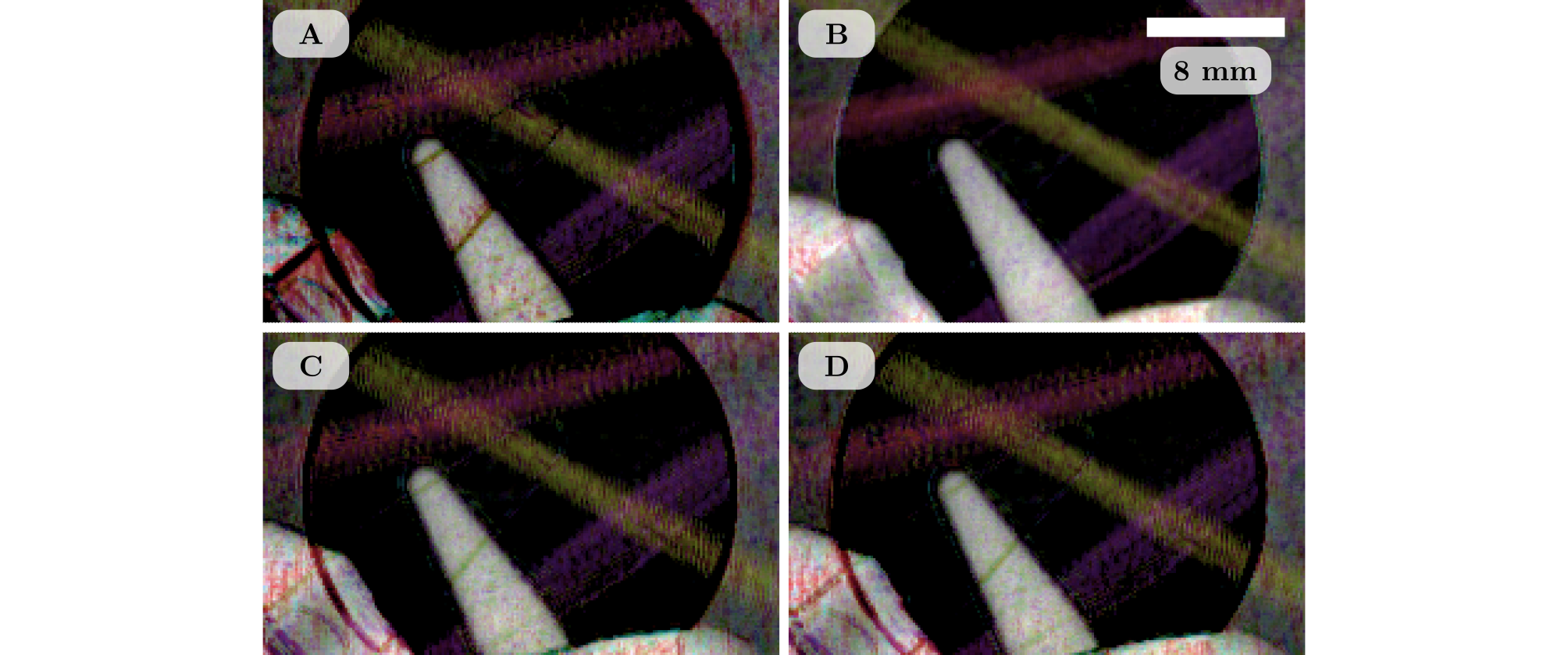}
\caption{Directional dark-field images of the dark-field phantom at 3.5~m sample-to-detector distance. Retrieved from uncorrected images (A), $I_s'$ removed using the experimental sample-only image (B), $I_s'$ removed using the transmission image retrieved from the single-grid directional dark-field retrieval algorithm (C), and $I_s'$ removed using an approximate sample-only image by scaling and dividing the sample-and-reference image by the reference-only image (D).}
\label{Fig: diff sample only}
\end{figure}

Figure~\ref{Fig: diff sample only}~A shows the directional dark-field image for the uncorrected images and Figure~\ref{Fig: diff sample only}~B shows the results for the experimental sample-only image divided data. These are the same images shown in Figure~\ref{Fig: corrections comparions}~C1 and~C2, assembled here for easy comparison with the approaches described in the previous paragraph. Figure~\ref{Fig: diff sample only}~C uses the retrieved sample transmission image as the approximate sample-only image while Figure~\ref{Fig: diff sample only}~D uses the scaling method to create the approximate sample-only image. Both approaches demonstrate similar improvements where the attenuation and propagation-base phase-contrast artefacts are reduced in comparison to the uncorrected directional dark-field image. The artefacts seen in Figure~\ref{Fig: diff sample only}~A are still partially present in Figure~\ref{Fig: diff sample only}~C and D. This is due to the approximate sample-only images having a lower spatial resolution and hence not containing the higher spatial frequency intensity changes that the experimental sample-only image captures. For some dark-field imaging applications, it is not possible to collect a sample-only image. In these cases, either of the above methods for creating an approximate sample-only image can be used to reduce artefacts.

\subsection*{Improvement in directional dark-field retrieval with lens blurring correction}
The flat-field and dark-current corrected reference-only image collected for the resin sample at the SPring-8 20XU beamline was used to measure the non-dark-field blurring due to the optical lens used in the detector. The lens locally blurs the image radially from the centre of the lens, with this blurring becoming stronger the further from the centre. The part of the reference-only image that had the most in-focus grid pattern was used as a reference region to compare to the rest of the image. The same cross-correlation analysis used in the single-grid directional dark-field retrieval algorithm~\cite{croughanDirectionalDarkfieldRetrieval2023} was utilised to measure the non-dark-field blurring kernels. The resin sample data was flat-field and dark-current corrected and rotated 3.5~degrees clockwise. The reference-only and sample-and-reference images were deblurred using Equation~\ref{Eq: deblurring operator} with the non-dark-field blurring kernels of $20 \times 20$ pixels in size for 10~iterations. A Fourier transform was performed on the reference-only image to determine the grid period to be 5.49~pixels and a local analysis kernel size of $9 \times 9$ pixels was used. The directional dark-field images were retrieved for the data with and without deblurring. 

\begin{figure}[ht]
\centering
\includegraphics[width=\linewidth]{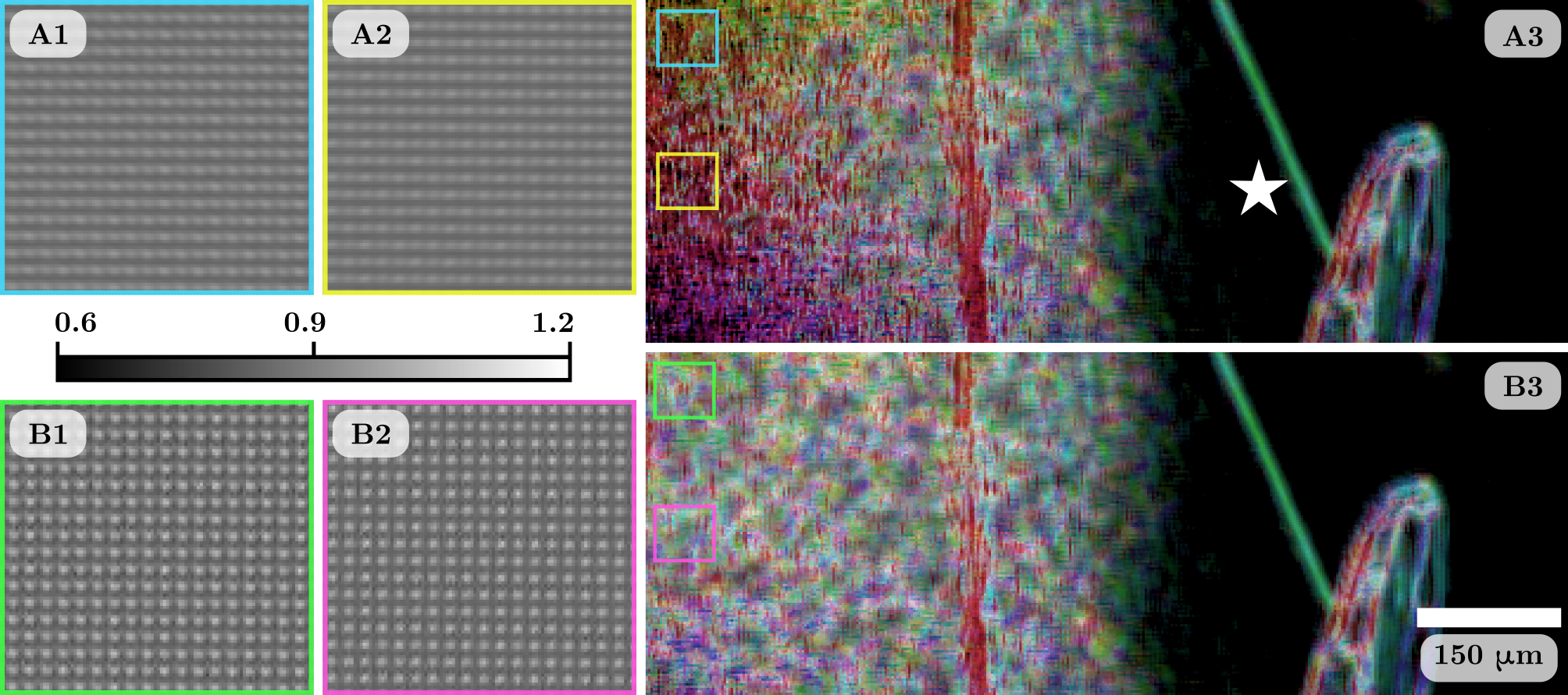}
\caption{Directional dark-field images of the resin sample for uncorrected data~(A3) and deblurred data~(B3) showing the cornflour in the left portion of the image and seashell calcium carbonate flakes in the middle and right of the image. The centre of the optical lens is shown with the white star in~A3. A1~and~A2 show the uncorrected reference-only image for the blue and yellow boxes in~A3, while~B1 and~B2 show the deblurred reference-only image for the green and pink boxes in~B3. See Figure~\ref{Fig: corrections comparions} for directional dark-field image scale bar, this time with a maximum $\Theta_{RMS}$~of~2.2~$\upmu$rad.}
\label{Fig: spring-8}
\end{figure}

Figure~\ref{Fig: spring-8}~A1 and~A2 show magnified sections of the uncorrected reference-only image demonstrating the difference in the reference pattern due to the position-dependent lens blur. Figure~\ref{Fig: spring-8}~B1 and~B2 show the same regions for the deblurred reference-only image. The visibility of the horizontal and vertical grid lines is increased and more uniform in the deblurred images. Figure~\ref{Fig: spring-8}~A3 and~B3 show the directional dark-field images for the uncorrected and deblurred data. The directional dark-field images have been cropped to zoom onto a section of the image with the centre of the lens (where the reference-only image was most in focus) marked by the white star in A3. The left half of~A3 and~B3 is where the cornflour is mixed in with the resin. The dark-field signal from the cornflour is expected to be isotropic and mostly consistent across the whole image. The lens blurring in the uncorrected data affects the measurement of the directional dark-field parameters. This results in artefacts in the left side of~A3 where the dark-field signal of the cornflour becomes a strong directional signal shown by the yellow, red, and pink hues. The dark-field blurring from the cornflour is strong enough that it almost completely blurs out the reference pattern in the sample-and-reference image. As the lens blurring becomes stronger in a particular direction, the visibility of the reference pattern in that direction becomes lower. The directional dark-field retrieval will only be sensitive in a direction up to the point where the changes in the reference pattern are detectable above the image noise. This means the maximum measurable blurring width is lower in the direction of the lens blurring than in the perpendicular direction. Thus strong isotropic dark-field blurring will appear to have a larger blurring width in the direction perpendicular to lens blur. The difference between the measured semi-major and semi-minor blurring widths will become larger as the lens blurring becomes stronger and the retrieved directional dark-field signal will become more asymmetrical. Deblurring the image data restores a uniform reference pattern with equal visibility in all directions. The directional dark-field image of this corrected data in Figure~\ref{Fig: spring-8}~B3 shows consistency in the dark-field signal retrieved from the cornflour across the whole image.

Strongly scattering samples that approach the saturation of a reference pattern that is asymmetrically blurred by non-dark-field sources can have an incorrect asymmetry measured due to the different visibilities in different directions in the reference pattern. This can result in the directional dark-field parameters being retrieved incorrectly and the directional dark-field image showing incorrect directional information. Deblurring the images removes this visibility differential and makes the retrieved results more consistent across the whole image. Utilising the position-dependent deblurring operator to remove lens blur can be beneficial in other types of imaging techniques.

\section*{Conclusion}

The x-ray dark-field reveals sub-pixel microstructures contained within a sample. Measurements of dark-field parameters such as the blurring widths and dominant scattering direction using the single-grid technique can be affected by non-dark-field experimental blur as well as attenuation and propagation-based phase contrast. It is expected that the dark-field blurring widths should increase proportionally with sample-to-detector distance and that the scattering angles should remain constant, however, it was previously found this was not the case~\cite{croughanDirectionalDarkfieldRetrieval2023}. In this paper we proposed a position-dependent blurring operator (Equation~\ref{Eq: blurring operator definition}) to model experimental blurring and presented an iterative inverse operation (Equation~\ref{Eq: deblurring operator}) to deblur experimental images. We also use a sample-only image, or, if not available, an approximation for a sample-only image, to remove the attenuation and propagation-based phase contrast. 

Data of a dark-field phantom was collected at the Imaging and Medical beamline at the Australian Synchrotron. This was imaged at nine sample-to-detector distances. By measuring the source-size blurring and detector point-spread-function, and capturing a sample-only image, we were able to deblur and remove the attenuation and propagation-based phase contrast to improve the directional dark-field image (Figure~\ref{Fig: corrections comparions}) and improve consistency of the measured blurring width as a function of sample-to-detector distance (Figure~\ref{Fig: distances analysis}). The edge artefacts present in the uncorrected directional dark-field image were successfully minimised by dividing the sample-and-reference image by a sample-only image. We proposed two approaches to create an approximate sample-only image and demonstrate that these approximate images can reduce the appearance of some artefacts (Figure~\ref{Fig: diff sample only}). This allows the artefacts to be partially corrected in data sets where collecting an experimental sample-only image is not possible. 

Finally, we showed a data set from the 20XU beamline at SPring-8 Japan, of a resin sample containing cornflour that produces a strong dark-field signal. The detector used had lens blur that increased in strength towards the edge of the image and resulted in dark-field artefacts where the dark-field signal saturates at a smaller blurring width in the direction of the lens blur. Deblurring the lens blur from the experimental images removed these artefacts from the retrieved directional dark-field image (Figure~\ref{Fig: spring-8}).

The corrections present in this paper have been shown to improve the measurement of the directional dark field using single-grid imaging. As these corrections are applied to the experimental images prior to dark-field retrieval we believe they have the potential to also improve other dark-field imaging techniques in similar ways. Further, the position-dependent deblurring operator can be utilised in other x-ray imaging approaches to remove unwanted experimental blur. In general, this correction can improve the spatial resolution and uniformity of spatial resolution of an x-ray image that is affected by experimental blur. This position-dependent deblurring correction can especially benefit x-ray imaging techniques that measure small spatial variations in intensity, which are most strongly affected by experimental blur. Phase imaging techniques such as single-grid phase-contrast imaging~\cite{wen2010single,morganSensitiveXrayPhase2013}, speckle-based imaging~\cite{zdoraStateArtXray2018} or propagation-based phase contrast~\cite{suzukiXrayRefractionenhancedImaging2002} can benefit from position-dependent deblurring, since this will improve the uniformity of the grid pattern, speckle pattern or propagation-based phase fringes across the image to allow for optimum phase retrieval. This can also be a beneficial correction for computed tomography, since sample features rotate across the image and so will experience varying degrees of blurring in each projection depending upon experimental blurring present at that position on the camera.

\section*{Acknowledgements}

The authors are grateful for the beamtime provided at the Imaging and Medical beamline at the Australian Synchrotron, part of ANSTO (Australia), under proposal 20055, and for the beamtime provided at the 20XU beamline at the SPring-8 Synchrotron, part of JASRI (Japan), under proposal number 2022B1169-BL20XU. K. S. M. acknowledges funding from the Australian Research Council (FT18010037, DP230101327). M. K. C., J. N. A., Y. Y. H. acknowledge funding from the Australian Research Training Program (RTP) and S.J. A. acknowledges a doctoral scholarship from the University of Canterbury. This research was supported by an AINSE Ltd. Postgraduate Research Award (PGRA) and the New Zealand Synchrotron Group Limited's capability fund grant.

\section*{Author contributions statement}
M. K. C. and Y. Y. H. conceived experiments. M. K. C., D. M. P., S. J. A, J. N. A., Y. Y. H., S. A. H., and K. S. M. conducted the experiments. M. K. C., D. M. P. and K. S. M., developed the analysis methods. M. K. C. analysed the results and drafted the manuscript. M. K. C., D. M. P., S. J. A, J. N. A., Y. Y. H., S. A. H., and K. S. M. reviewed the manuscript. 

\section*{Data Availability Statement}
The datasets generated during and/or analysed during the current study are available from the corresponding author on reasonable request.
\bibliographystyle{unsrt}  
\bibliography{Manuscript}

\end{document}